\documentclass[12pt,letterpaper]{article}
\usepackage{jheppub}
\usepackage{amsmath}
\DeclareMathOperator{\Tr}{Tr}

\title{Proof of a Quantum Bousso Bound }

\author[a,b]{Raphael Bousso,}
\author[c,d]{Horacio Casini,}
\author[a,b]{Zachary Fisher,}
\author[d]{and Juan Maldacena}
\affiliation[a]{Center for Theoretical Physics and Department of Physics,\\
 University of California, Berkeley, CA 94720, U.S.A.}
\affiliation[b]{Lawrence Berkeley National Laboratory, Berkeley, CA 94720,
  U.S.A.}
\affiliation[c]{Centro At\'omico Bariloche, 8400 Bariloche, R\'{\i}o Negro, Argentina}
\affiliation[d]{Institute for Advanced Study, Princeton, NJ 08540, USA}

\abstract{We prove the generalized Covariant Entropy Bound, $\Delta S\leq (A-A')/4G\hbar$, for light-sheets with initial area $A$ and final area $A'$.  The entropy $\Delta S$ is defined as a difference of von Neumann entropies of an arbitrary state and the vacuum, with both states restricted to the light-sheet under consideration.  The proof applies to free fields, in the limit where gravitational backreaction is small.
 We do not assume the null energy condition. In regions where it is violated, we find that the bound is protected by the defining property of light-sheets: that their null generators are nowhere expanding.  }

\begin{document}
\maketitle

\section{Introduction}
\label{sec-intro}

The study of black hole thermodynamics has led to some interesting entropy bounds that should be obeyed for the consistency of the theory.  The simplest one is the Bekenstein bound, which does not involve Newton's constant \cite{Bek81}. This bound, when properly formulated, \cite{Cas08} (see also \cite{BlaCas13}), is a simple consequence of relativistic quantum field theory. 

A different kind of bound involves bounding entropies by areas in Planck units.  These bounds are inspired by the black hole entropy formula. The most general bound of this type is the Bousso bound \cite{CEB1} or {\em Covariant Entropy Bound}. It can be applied not only to matter crossing black hole horizons, but also to rapidly expanding or collapsing regions that cannot be converted to black holes.  Thus it transcends the original motivation from black hole thermodynamics.\footnote{A key motivation was the holographic principle~\cite{Tho93,Sus95}. Fischler and Susskind~\cite{FisSus98} pioneered the search for a holographic entropy bound in cosmology.}

The Covariant Entropy Bound states that the entropy $\Delta S$ on a {\em light-sheet} cannot exceed its initial area $A$:
\begin{equation}
 { A \over 4 G \hbar }  \geq \Delta S \,,
\label{eq-ceb}
\end{equation}
A light-sheet is a null hypersurface whose cross-sectional area is decreasing or staying constant, in the direction away from $A$. 

A light-sheet can be constructed by starting with any surface $A$,\footnote{$A$ must be spacelike and of codimension two in the spacetime. It need not be closed. We use $A$ to denote both the surface and its area.} in any spacetime. There are four orthogonal null directions, past- and future-directed to either side of $A$. A light-sheet is generated by null geodesics that have nonpositive expansion, $\theta\leq 0$, away from $A$. This is a local condition and it is required to hold at every point on the light-sheet. When it breaks down, e.g.\ at caustics where neighboring generators intersect, the corresponding generator must be terminated. If $A$ has more than one light-sheet, the bound can be applied to each individually.

If any generators are terminated before a caustic is reached, then the cross-sectional area $A'$ of the endpoints of the light-sheet will not vanish. In this case the conjecture can be strengthened~\cite{FlaMar99}:
\begin{equation}
 { \Delta A \over 4 G \hbar }  \geq \Delta S \,.
\label{eq-gceb}
\end{equation}
The difference between the initial and final area, $\Delta A=A-A'$, is nonnegative because the expansion $\theta\leq 0$ is the logarithmic derivative of the area transverse to the null generators, with respect to an affine parameter that increases away from $A$~\cite{Wald}. 

Fundamentally, the Covariant Entropy Bound is a conjecture. It might capture aspects of how spacetime and matter arise from a more fundamental theory~\cite{CEB2,RMP}. A general proof may not become available until such a theory is found. Nevertheless, it is of interest to prove the bound at least in certain regimes, or subject to assumptions that hold in a large class of examples. 

In this spirit, the bound (\ref{eq-gceb}) has been shown to hold in settings where the entropy $\Delta S$ can be approximated hydrodynamically, as the integral of an entropy flux over the light-sheet; and where suitable relations constrain the entropy and energy fluxes~\cite{FlaMar99,BouFla03}. These assumptions apply to a large class of spacetimes, such as cosmology or the gravitational collapse of a star. Thus they establish the broad validity of the bound. But the underlying assumptions have no fundamental status, for two reasons that we will now describe.

Unlike the stress tensor, entropy is not local, so the hydrodynamic approximation breaks down if the light-sheet is shorter than the modes that dominate the entropy. In this regime, it is not clear how to define the entropy at all. Consider a single photon wavepacket with a Gaussian profile propagating through otherwise empty flat space. In order to obtain the tightest bound, we may take the light-sheet to have initially vanishing expansion. $\Delta A$ is easily computed from the stress tensor and Einstein's equations. For a finite light-sheet that captures all but the exponential tails of the wavepacket, one finds that the packet focuses the geodesics just enough to lose about one Planck area, $\Delta A/G\hbar\sim O(1)$~\cite{Bou03}. For smaller light-sheets, $\Delta A$ tends to 0 quadratically with the affine length. For larger light-sheets, $\Delta A$ can grow without bound. To check if the bound is satisfied for all choices of light-sheet, one would need a formula for the entropy on any finite light-sheet. Globally, the entropy is $\log n\sim O(1)$, where $n$ is the number of polarization states. Intuitively this should also be the answer when nearly all of the wavepacket is captured on the light-sheet, but how can this be quantified? (In field theory, the entropy in a finite region would be dominated by vacuum entanglement entropy across the initial and final surface, and hence largely unrelated to the photon.)  Worse, for short light-sheets, there is no intuitive notion of entropy at all. What is the entropy of, say, a tenth of a wavepacket?\footnote{Similar limitations apply to the Bekenstein bound~\cite{Bek81}, which can be recovered as a special case of the generalized covariant bound in the weak-gravity limit~\cite{Bou03}: precisely in the regime where the bound becomes tight, one lacks a sharp definition of entropy.}

A second limitation of the sufficient conditions identified in Refs.~\cite{FlaMar99,BouFla03} is that the assumed inequalities between entropy and energy flux imply the null energy condition. This condition on the stress tensor does not hold in all regions for all quantum states. Hence, independently of the validity of the hydrodynamic limit, the sufficient conditions of Refs.~\cite{FlaMar99,BouFla03} need not hold. An example of a region where the null energy condition is violated is the horizon of an evaporating black hole. Indeed, it has been argued \cite{Low99,StrTho03} that by critically illuminating a black hole so as to keep its horizon area constant, an arbitrary amount of entropy can be passed through a light-sheet. This violates the bound (\ref{eq-ceb}) only over a timescale on which quantum corrections to the geometry become dominant. However, the stronger bound (\ref{eq-gceb}) becomes violated immediately, and thus in a regime where the gravitational backreaction from both Hawking radiation and infalling matter is small.

In this article we will address the above difficulties for the case that matter consists of free fields, and in the limit of weak gravitational backreaction. We will provide a sharp definition of the entropy on a finite light-sheet in terms of differences of von Neumann entropies. Our definition does not rely on a hydrodynamic approximation. It reduces to the expected entropy flux in obvious settings. Using this definition, we will prove the covariant bound. We will not assume the null energy condition.


\paragraph{Outline} In Sec.~\ref{sec-def} we provide a definition of the entropy on a weakly focused light-sheet. We define $\Delta S$ as the difference between the entropy of the matter state and the entropy of the vacuum, as seen by the algebra of operators defined on the light-sheet.

The proof of the bound then has two steps.  In Sec.~\ref{sec-SK}, we note that $\Delta S \leq \Delta K$, where $\Delta K$ is the difference in expectation values for the vacuum modular Hamiltonian. This property holds for general quantum theories~\cite{Cas08}. In Sec.~\ref{sec-KA}, we show that $\Delta K\leq \Delta A/4G\hbar$. We first compute an explicit expression for the modular Hamiltonian, in Sec.~\ref{sec-Kcomp}. For general regions, the modular Hamiltonian is complicated and non-local. However, the special properties of free fields on light-like surfaces enable us to derive explicitly the modular Hamiltonian in terms of the stress tensor.  The expression is essentially the same as the result we would obtain for a null interval in a 1+1 dimensional CFT.  Finally, in Sec.~\ref{sec-acomp}, we use the Raychaudhuri equation to compute the area difference $\Delta A$. The area difference comes from two contributions: focusing of light-rays by matter, and potentially, a strictly negative initial expansion. Usually one may choose the initial expansion to vanish. If this choice is possible, it will minimize $\Delta A$ and provide the tightest bound. However, if the null energy condition is violated, it can become necessary to choose a negative initial expansion, in order to keep the expansion nonpositive along the entire interval in question and evade premature termination of the light-sheet. We find that the two contributions together ensure that $\Delta A/4G\hbar \geq \Delta K$. Combining the two inequalities, we obtain the covariant bound, $\Delta A/4 G \hbar  \geq \Delta S $.

In Sec.~\ref{sec-discussion}, we discuss possible generalizations of our result to the cases of interacting fields and large backreaction.  We comment on the relation of our work to Casini's proof of Bekenstein's bound from the positivity of relative entropy~\cite{Cas08}, to Wall's proof of the generalized second law~\cite{Wal11}, and to an earlier proposal for incorporating quantum effects in the Bousso bound~\cite{StrTho03}.

In the Appendix, we prove monotonicity of $\Delta A/(4G\hbar)-\Delta S$ under inclusion, a result stronger than that obtained in the main body of the paper.

\section{Regulated Entropy $\Delta S$}
\label{sec-def}

We will consider matter in asymptotically flat space, perturbatively in $G$. Since Minkowski space is a good approximation to any spacetime at sufficiently short distances, our final result should apply in arbitrary spacetimes, if the transverse and longitudinal size of the lightsheet is small compared to curvature invariants. For definiteness, we work in 3+1 spacetime dimensions; the generalization to $d+1$ dimensions is trivial.

\begin{figure}[ht!]
\begin{center}
\includegraphics[scale=0.6]{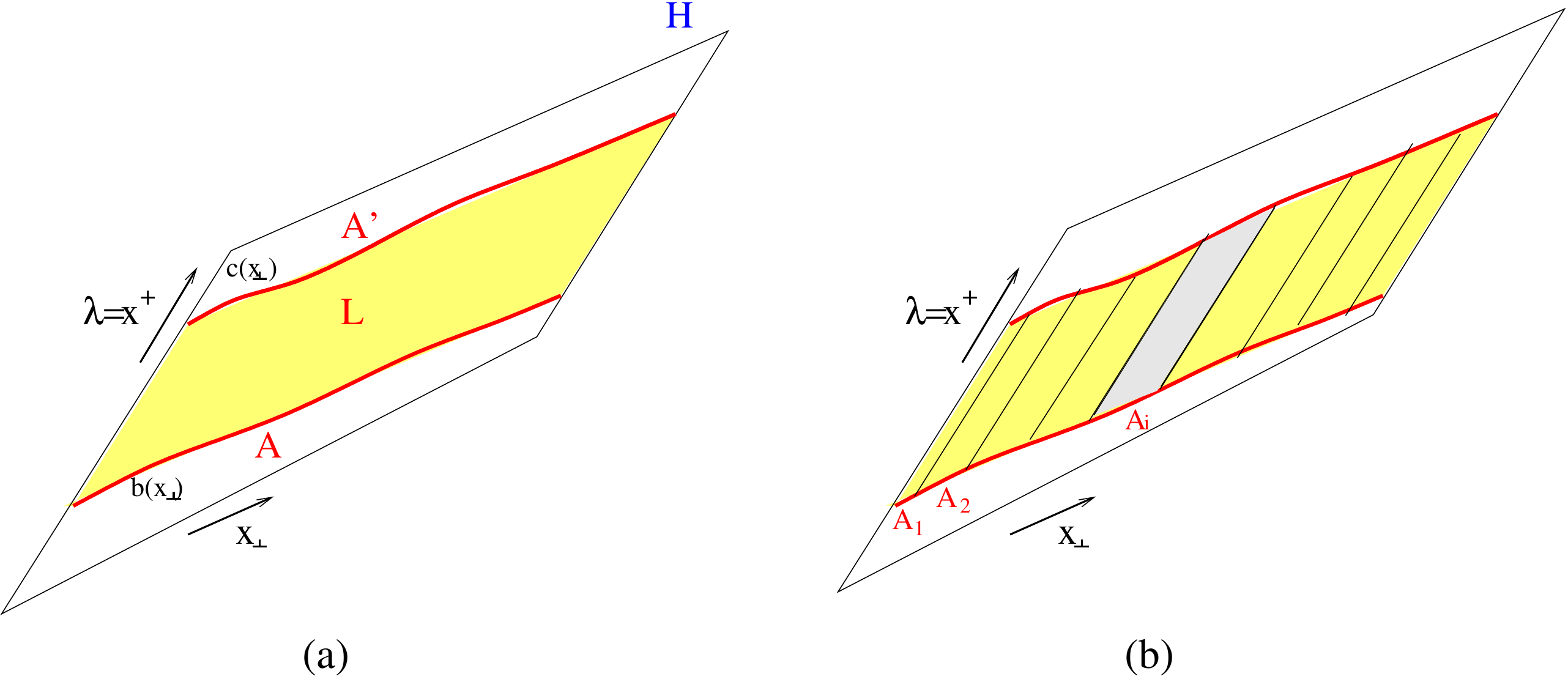}
\caption{The light-sheet $L$ is a subset of the light-front $x^-=0$, consisting of points with $b(x_\perp ) \leq x^+ \leq c(x_\perp )$ (a). The light-sheet can be viewed as the disjoint union of small transverse neighborhoods of its null generators (b).}
\label{lightsheets}
\end{center}
\end{figure}

At zeroth order in $G$, the metric is that of Minkowski space:
\begin{equation}
ds^2 = -dx^+ dx^- + dx_\perp^2~,
\end{equation}
where $dx_\perp^2 = dy^2+dz^2$. Without loss of generality, we will consider a partial light-sheet $L$ that is a subset of the null hypersurface $H$ given by $x^-=0$. Any such light-sheet can be characterized by two piecewise continuous functions $b(x_\perp)$ and $c(x_\perp)$ with $-\infty< b\leq c<\infty$ everywhere: $L$ is the set of points that satisfy $x^-=0$, $b<x^+<c$. See figure \ref{lightsheets}. 

We begin by giving an intrinsic definition of the vacuum state on $H$ in free field theory. The generator of a null translation $x^+ \to x^+ + a(x_\perp)$ along $H$ is given by
\begin{equation}
p_+[a] = \int dx^2_\perp \int_{-\infty}^\infty dx^+ T_{++}\, a(x_\perp)~,
\label{eq-pa}
\end{equation}
where $T_{++}=T_{ab} k^a k^b$ and $k^a= \partial_{+}$ is the tangent vector to $H$. 
Given any choice of $a(x_\perp)$, one can define a vacuum state $|0\rangle_a$ by the condition $p_+[a] |0\rangle _a=0$.

In fact, all nowhere vanishing functions $a(x_\perp)$ define the same vacuum, $|0\rangle_H$, because of the following important result~\cite{Wal11}: there are neither interactions nor correlations\footnote{These statements hold for correlators that have at least one derivative along the plus direction $\partial_+ \phi$. Correlators of $\phi$ with no derivatives are non-zero 
at spacelike distances. However, they do not lead to well defined operators along the light front since we cannot control the UV divergences by smearing it along the light front directions. For this reason we do not consider $\phi$ as part of the algebra ${\cal A}(H)$. The canonical stress tensor component $T_{++} \propto (\partial_+ \phi)^2 $ depends only on such  derivatives of the field in the null direction. For further details, see Ref.~\cite{Wal11}.} between different null generators of $H$. When restricted to $H$, the algebra of observables ${\cal A}$ becomes {\em ultralocal\/} in the transverse direction. For any partition $\{H_i\}$ of the null generators of $H$, the algebra can be written as a tensor product
\begin{equation}
{\cal A}(H) = \prod_i {\cal A}(H_i)~.
\label{eq-ul}
\end{equation} 
In the limit where the translation is localized to one ray, $a(x_\perp') = \delta(x_\perp'-x_\perp)$, Eq.~(\ref{eq-pa}) reduces to the generator
\begin{equation}
p_+(x_\perp) = \int_{-\infty}^\infty dx^+ T_{++} ~,
\end{equation}
and $p_+(x_\perp)|0\rangle_{x_\perp}=0$ defines a vacuum state independently for each generator. By ultralocality, the vacuum state on $H$ is a tensor product of these states. (In terms of small transverse neighborhoods of each generator, $H_i$, one can write $|0\rangle_H = \prod_i |0\rangle _i$.) 

It will be convenient to write the vacuum state on $H$ as a density operator,
\begin{equation}
\sigma_H \equiv |0\rangle_H \mbox{}_H\langle 0| \,.
\end{equation}
Let the actual state of matter on $H$ be $\rho_H$; this state may be mixed or pure. Let $\sigma_L$ and $\rho_L$ be the restriction, respectively, of the vacuum and the actual state to the lightsheet $L$:
\begin{eqnarray}
\sigma_L & \equiv & \Tr_{H-L} \sigma_H \\
\rho_L & \equiv & \Tr_{H-L} \rho_H
\end{eqnarray}
The von Neumann entropy of either of these density matrices diverges in proportion to the sum of the areas of the two boundaries of $L$ (in units of a UV cutoff). However, we may define a \textit{regulated entropy} as the difference between the von Neumann entropies of the actual state and the vacuum~\cite{Cas08,MarMin04,HolLar94}:
\begin{equation} \label{EntropyDifference}
\Delta S \equiv S(\rho_L) - S(\sigma_L) = - \Tr\rho_L\log\rho_L + \Tr\sigma_L\log\sigma_L~.
\end{equation}
For finite energy global states $\rho_H$, this expression will be finite and independent of the regularization scheme. It reduces to the global entropy, 
$\Delta S\to -\Tr \rho_H\log\rho_H$, in the limit where the latter is dominated by modes that are well-localized to $L$. Examples include large thermodynamic systems such as a bucket of water or a star, but also a single particle wavepacket that is well-localized to the interior of $L$.

\begin{figure}[ht!]
\begin{center}
\includegraphics[scale=1.2]{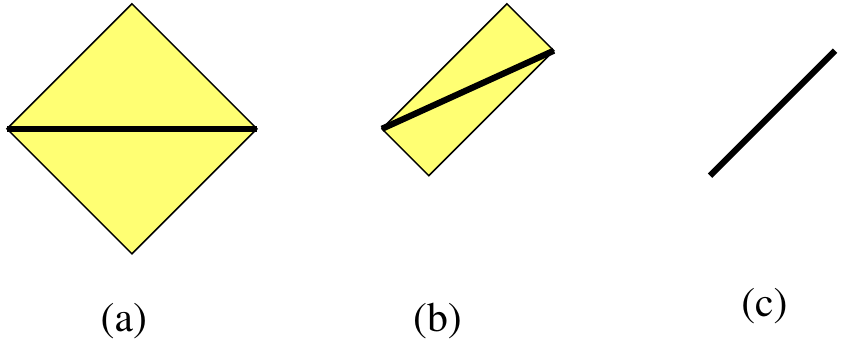}
\caption{Operator algebras associated to various regions. (a) Operator algebra associated to the domain of dependence (yellow) of a spacelike interval. (b) The domain of dependence of a boosted interval. (c) In the null limit, the domain of dependence degenerates to the interval itself.}
\label{boostedintervals}
\end{center}
\end{figure}
An important feature is that we are computing these entropies for null segments.  It is more common to consider entropies for spatial segments, see figure \ref{boostedintervals}. In that case, the algebra of operators includes all the local operators in the domain of dependence of the segment, see figure \ref{boostedintervals}(a). We can also consider a boosted the interval as in figure \ref{boostedintervals}(b).  The domain of dependence changes accordingly. In the limit of a null interval the domain of dependence becomes just a null segment. This is a singular limit of the standard spacelike case: the proper length of the null interval vanishes and the domain of dependence degenerates. Despite these issues, we find that the entropy difference between any state and the vacuum, (\ref{EntropyDifference}), is finite and well defined. In the free theory case, the limiting operator algebra has the ultralocal structure described above.

\section{Proof that $\Delta S\leq \Delta K$}
\label{sec-SK}

The vacuum state on the light-sheet $L$ defines a \textit{modular Hamiltonian} operator $K_L$, via
\begin{equation}
\sigma_L = \frac{e^{-K_L}}{\Tr e^{-K_L}}~,
\label{eq-kldef}
\end{equation}
up to a constant shift that drops out below. Expectation values such as $\Tr K_L\sigma_L$ and $\Tr K_L \rho_L$ will diverge, but we may define a regulated  (or vacuum-subtracted) 
\textit{modular energy} of $\rho_L$:
\begin{equation}
\Delta K \equiv \Tr K_L\rho_L - \Tr K_L \sigma_L~.
\label{eq-kdef}
\end{equation}
 
For any two quantum states $\rho,\sigma$, in an arbitrary setting, one can show that the \textit{relative entropy},
\begin{equation}
S(\rho|\sigma) \equiv \Tr\rho\log\rho-\Tr\rho\log\sigma~,
\end{equation}
is nonnegative~\cite{Lin73}.\footnote{Moreover, the relative entropy decreases monotonically under restrictions of $\rho, \sigma$ to a subalgebra~\cite{Lin75}. With the help of this stronger property, our conclusion can be strengthened to the statement that $\frac{\Delta A(c,b)}{4G\hbar}-\Delta S$ decreases monotonically to zero if the boundaries $b$ and $c$ are moved towards each other. This is shown in the Appendix.} With the above definitions, this immediately implies the inequality~\cite{Cas08}
\begin{equation}
\Delta S\leq \Delta K~.
\end{equation}
To prove the generalized Covariant Entropy Bound, we will now show that $\Delta K\leq \Delta A/4G\hbar$, where $\Delta A$ is the area difference between the two boundaries of the light-sheet.

\section{Proof that $\Delta K\leq \Delta A/4G\hbar$}
\label{sec-KA}

We can think of the null hypersurface $H$ as the disjoint union of small neighborhoods $H_i$ of a large discrete set of null generators; see figure \ref{lightsheets}(b). By ultralocality of the operator algebra, Eq.~(\ref{eq-ul}), we have for the vacuum state $\sigma_H = \prod_i \sigma_{L,i}$, $\sigma_L = \prod_i \sigma_{L,i}$, where the density operators for neighborhood $i$ are defined by tracing over all other neighborhoods~\cite{Wal11}. Using $\sigma_i$ in Eqs.~(\ref{eq-kldef}) and (\ref{eq-kdef}), a modular energy $\Delta K_i$ can be defined for each neighborhood, which is additive by ultralocality: $\Delta K = \sum_i \Delta K_i$. Strictly, we should take the limit as the cross-sectional area of each neighborhood becomes the infinitesimal area element orthogonal to each light-ray, $A_i\to d^2 x_\perp$. However, we find it more convenient to think of $A_i$ as finite but small, compared to the scale on which the light-sheet boundaries $b$ and $c$ vary.

Since both the modular energy and the area are additive,\footnote{By contrast, the entropy $\Delta S$ is {\em subadditive} over the transverse neighborhoods. In Eq.~(\ref{EntropyDifference}), the vacuum state $\sigma_L$ factorizes, but the general state $\rho_L$ can have entanglement across different neighborhoods $H_i$. This does not affect our argument since we have already shown directly that $\Delta S \leq \Delta K$.} it will be sufficient to show that $\Delta K_i\leq \Delta A_i/4G\hbar$, where $\Delta A_i$ is the change in the cross-sectional area $A_i$ produced at first order in $G\hbar$ by gravitational focusing. We will demonstrate this by evaluating $\Delta K_i$ and bounding $\Delta A_i$. For any given neighborhood $H_i$, we may take the affine parameter $\lambda_i$ to run from $0$ to $1$ on the light-sheet $L_i$, as $x_+$ runs from $b_i=b(x_\perp)$ to $c_i=c(x_\perp)$. 



For notational simplicity we will drop the index $i$ in the remainder of this section. 

\subsection{Ultralocality and Conformal Symmetry Determine $\Delta K$}
\label{sec-Kcomp}

We compute the modular Hamiltonian $K_L$ on the null interval $0< \lambda < 1$ in two steps. First, we review the modular Hamiltonian for the semi-infinite interval $1<\lambda'<\infty$. Then we use the special conformal symmetry of the algebra of observables ${\cal A}$ to obtain $K_L$ by inversion.

We can regard the interval $1<\lambda'<\infty$ as the upper boundary of a right Rindler wedge with bifurcation surface $\lambda'=1$. By tracing the global vacuum $\sigma$ over the left Rindler wedge, one finds that the state on the right is given by the thermal density operator
\begin{equation}
\sigma_{RW} = \frac{e^{-K_{RW}}}{\Tr e^{-K_{RW}}}~,
\end{equation}
where the modular Hamiltonian
\begin{equation}
K_{RW} = \frac{2\pi}{\hbar} \int d^2x_\perp \int_1^\infty d\lambda'\, (\lambda' -1)\, T_{\lambda'\lambda'}
\end{equation}
coincides with the well-known Rindler Hamiltonian.

Wall~\cite{Wal11} has shown that the horizon algebra on each generator of $H$ is that of the left-moving modes of a 1+1 dimensional conformal field theory. General states transform nontrivially, but the vacuum $\sigma$ is invariant under special conformal transformations. Hence, the modular Hamiltonian on the interval $0<\lambda<1$ can be obtained by applying an inversion $\lambda'\to \lambda=1/\lambda'$ to the Rindler Hamiltonian. Using
\begin{equation}
T_{\lambda'\lambda'} = T_{\lambda\lambda} \left(\frac{d\lambda}{d\lambda'}\right)^2~,
\end{equation}
one finds for the modular Hamiltonian of the light-sheet $L$:
\begin{equation}
K_L = \frac{2\pi}{\hbar} \int d^2x_\perp \int_0^1 d\lambda\, \lambda (1-\lambda)\, T_{\lambda\lambda}~.
\label{eq-k}
\end{equation}

Let us make some comments. If we were dealing with a two dimensional CFT the formula (\ref{eq-k}) would be familiar.  If instead we had a massive free field in two dimensions, then we note that a null interval is conceptually similar to a very small interval.  Therefore we are exploring the UV properties of the theory, which are the same as those for a massless free field. When we go to higher dimensions we can understand (\ref{eq-k}) as the result of thinking of the free field in terms of a two dimensional massive fields with masses given by a Kaluza Klein reduction along the transverse dimensions.

\subsection{Focusing and Nonexpansion Bound $\Delta A$}
\label{sec-acomp}

Generally, the expansion of a null congruence is defined as \cite{Wald}
\begin{equation}
\theta(\lambda) \equiv \widehat{\nabla_a k^a} = \frac{d \log \delta\! A}{d\lambda}
\label{eq-thetadef}
\end{equation}
where $\delta\! A$ is an infinitesimal cross-sectional area element. Recall that in the present context we consider the transverse neighborhood of one null geodesic, with small cross section $A_i$, so we may replace $\delta\! A\approx A_i$. Our task is to compute the {\em change\/} $\Delta A_i$ of this small cross-section, from one end of $L_i$ to the other, by integrating Eq.~(\ref{eq-thetadef}). We will drop the index $i$, as it suffices to consider any one neighborhood.

At zeroth order in $G\hbar$, the light-sheet of interest is a subset of the null plane $x^-=0$ in Minkowski space, and so has vanishing expansion $\theta$ and vanishing shear $\sigma_{ab}$ everywhere. One may compute the expansion at first order in $G\hbar$ by integrating the Raychaudhuri equation
\begin{equation}
\frac{d\theta}{d\lambda} = -\frac{1}{2}\theta^2 -
\sigma_{ab}\sigma^{ab} - 8 \pi G T_{\lambda\lambda}~,
\label{eq-ray}
\end{equation}
The twist $\omega_{ab}$ vanishes identically for a surface-orthogonal congruence.

We will pick $\lambda=0$ as the initial surface and integrate up to $\lambda=1$. The choice of direction is nontrivial, since we must ensure that the defining condition of light-sheets is everywhere satisfied: the cross-sectional area must be nonexpanding away from the initial surface, everywhere on $L$. As we shall see, this implies that at first order in $G\hbar$, we must allow for a nonzero initial expansion $\theta_0$ at $\lambda=0$. The required initial expansion can be accomplished by a small deformation of the initial surface~\cite{Bou03}, whose effects on $\Delta K$ and $\Delta S$ only appear at higher order. (Of course, we could also start at $\lambda=1$ and integrate in the opposite direction. For any given state, both $\Delta A$ and the initial expansion will depend on the choice of direction. But we will demonstrate that $\Delta K\leq \Delta A$ for {\em all\/} states on future-directed light-sheets beginning at $\lambda=0$. By symmetry of $K_L$ under $\lambda\to 1-\lambda$, the same result immediately follows for past-directed light-sheets beginning at $\lambda=1$.)

From Eq.~(\ref{eq-ray}) we obtain at first order in $G\hbar$:
\begin{equation}
\theta(\lambda) = \theta_0 - 8 \pi G \int_0^\lambda T_{\hat\lambda\hat\lambda} d\hat\lambda ~.
\label{eq-thetalambda}
\end{equation}
The nonexpansion condition is
\begin{equation}
\theta(\lambda)\leq 0, ~~~~\mbox{for all~}\lambda\in [0,1]~.
\label{eq-nonexp}
\end{equation}
If the null energy condition holds, $ T_{\lambda\lambda}\geq 0$, then this condition reduces to $\theta_0\leq 0$. More generally, however, we may have to choose $\theta_0<0$ to ensure that antifocusing due to negative energy densities does not cause the expansion to become positive, and thus the light-sheet to terminate, before $\lambda=1$ is reached. However, it is always sufficient to take $\theta_0$ to be of order $G\hbar$, so it was self-consistent to drop the quadratic terms $\propto \theta^2$, $\sigma_{ab}\sigma^{ab}$, in the focusing equation. Note that, in the semiclassical 
quantization scheme,  the $\sigma^2$ term
can be viewed as arising from the stress tensor of the gravitons and can be explicitly included as part of the total stress tensor by separating the 
gravitational field into long and short distance modes.

From the definition of the expansion, Eq.~(\ref{eq-ray}), one obtains the difference between initial and final cross-sectional area:
\begin{equation}
\frac{\Delta A}{A} = - \int_0^1 d\lambda \theta(\lambda) = -\theta_0 + 8\pi G \int_0^1 d\lambda (1-\lambda) T_{\lambda\lambda}~,
\label{eq-inttheta}
\end{equation}
where we have used Eq.~(\ref{eq-thetalambda}) and exchanged the order of integration. In order to eliminate $\theta_0$ we now use the nonexpansion condition: let $F(\lambda)$ be a function obeying $F(0)=0$, $F(1)=1$ and $F'(\lambda) \geq 0$ for $ 0 \leq \lambda \leq 1 $. 
From Eq.~(\ref{eq-nonexp}), we have $0\geq \int_0^1 F' \theta d\lambda$, and thus from (\ref{eq-thetalambda}) and integration by parts we find
\begin{equation}
\theta_0 \leq 8\pi G \int d\lambda [1-F(\lambda)] T_{\lambda\lambda}~.
\label{eq-bth}
\end{equation}
With the specific choice $F(\lambda) = 2\lambda -\lambda^2$ we find from Eqs.~(\ref{eq-inttheta}) and (\ref{eq-bth}) that the area difference is bounded from below by the modular Hamiltonian:
\begin{equation}
\Delta A \geq A\times 8\pi G \int_0^1 d\lambda\, \lambda (1-\lambda) \,T_{\lambda\lambda}~.
\label{eq-intthetares}
\end{equation}
Comparison with Eq.~(\ref{eq-k}) shows that $\Delta K\leq \Delta A/4G\hbar$, as claimed.

Combined with the earlier result $\Delta S\leq \Delta K$, this completes the proof of the Covariant Entropy Bound, $\Delta S\leq \Delta A/4G\hbar$, for free fields in the weak gravity limit.

\section{Discussion}
\label{sec-discussion}

An interesting aspect of this argument is that we did not need to assume any microscopic relation between energy and entropy.  We did have to assume that we had a local quantum field theory at short distances. Therefore the necessary relation between entropy and energy is the one automatically present in quantum field theory, i.e., given by the explicit expression of the modular Hamiltonian in terms of the stress tensor. Our discussion required a careful definition of the entropy that appeared in the bound. In that sense it is very similar to the Casini version \cite{Cas08} of the Bekenstein bound (see also \cite{MarMin04,HolLar94}), and also to Wall's proof of the generalized second law~\cite{Wal10,Wal11}.

All these developments underscore the interesting interplay between local Lorentz invariance of the quantum field theory, Einstein's equations, and information.  It has often been speculated that the validity of these entropy bounds would require extra constraints on the matter that is coupled to Einstein's equations. Here we see that the only constraint is that matter obeys the standard rules of local quantum field theory. (Conversely, it may be possible to view these rules as a consequence of entropy bounds~\cite{Bou04}.)

\paragraph{Relation to other work}
In \cite{Low99} a possible counterexample to the Covariant Entropy Bound was proposed. The idea is to feed matter so slowly into an evaporating black hole that the horizon area remains static or slowly decreases during the process. Hence the horizon is a future-directed light-sheet, to which the bound applies. Yet, it would appear that one can pass a very large amount of entropy through the horizon in this way. How is this consistent with our proof? 

To understand this, consider the simplest case where the stress tensor component $T_{++}$ is constant on the light-sheet. For the horizon area to stay constant or shrink, one must have $T_{++}\leq 0$. By Eq.~(\ref{eq-k}), this implies $\Delta K\leq 0$,\footnote{We have considered the case where the light-sheet $L$ is a portion of a null plane $H$ in Minkowski space, whereas we are now discussing the case where $L$ is a portion of the horizon $H$ of a black hole. In general, application of our flat space results to general spacetimes would require that the transverse size of $L$ be small compared to the curvature scale. This is not the case for the horizon of a black hole. However, the vacuum states $\sigma_H$ and $\sigma_L$ can be defined directly on the black hole background; $\sigma_H$ is the Hartle-Hawking vacuum.} and positivity of the relative entropy requires $\Delta S\leq \Delta K$.  Hence, in this case, $\Delta S\leq 0$. Thus we find that with our definitions, the entropy is negative for an evaporating black hole, even with the addition of some positive, partially compensating flux; and the entropy is at least nonpositive in the static case. Since $\Delta A\geq 0$ by the nonexpansion condition, the bound is safe.

Strominger and Thompson \cite{StrTho03} have also proposed a quantum version of the Covariant Entropy Bound. Their proposal is analogous to the definition of generalized entropy, in that one adds to the area the entanglement entropy of quantum fields that are outside the horizon and distinct from the matter crossing the light-sheet. In contrast, we have given a definition which only involves properties of the quantum fields on the light-sheet $L$, i.e., on the relevant portion of the horizon.

A similar distinction must be made when comparing our result to Wall's proof of the generalized second law~\cite{Wal10,Wal11}. Wall considers the generalized entropy $S_{\rm gen}(A)=S_m(A)+A/4G\hbar$ on semi-infinite horizon regions, where $A$ the area of a horizon cross-section, and $S_m(A)$ is the matter entropy on the portion the horizon to the future of $A$ (which is closely related to the matter entropy on spatial slices exterior to $A$). Given two horizon slices with $A_2$ to the future of $A_1$, monotonicity of the relative entropy under restriction of the semi-infinite null hypersurface starting at $A_1$ to the semi-infinite subset starting at $A_2$ implies the GSL:
\begin{equation} 
0\leq S_{\rm gen}(A_1)-S_{\rm gen}(A_2)~.
\end{equation}
The argument applies to causal horizons, such as Rindler and black hole horizons.

Unlike our proof of the covariant bound, Wall's proof (like that of \cite{StrTho03}) does {\em not} assume the nonexpansion condition. This is as it should be, since the GSL does not require any such condition. Suppose, for example, that the expansion is not monotonic between $A_1$ and $A_2$, because the black hole is evaporating but there is also matter entering the black hole. Then the horizon interval from $A_1$ to $A_2$ is not a light-sheet with respect to either past- or future-directed light-rays. Yet, the GSL must hold. On the other hand, our proof applies to all weakly focussed null hypersurfaces, whereas the GSL applies only to causal horizons. 

Now suppose we consider a case where both the GSL and the covariant bound should apply, such as a monotonically shrinking or growing portion of a black hole horizon. In this case, it should be noted that our proof and Wall's proof \cite{Wal10,Wal11} refer to different entropies. In general the difference in the matter entropy outside $A_1$ and $A_2$ is distinct from the entropy that we have defined directly on the interval stretching from $A_1$ to $A_2$: 
\begin{equation}
{\cal D}S\equiv S_m(A_1)-S_m(A_2)\neq \Delta S~.
\end{equation}
Because ${\cal D}S-\Delta S$ is not of definite sign (and because of the different assumptions about nonexpansion), our result does not imply Wall's, and his does not imply ours even in the special case where a horizon segment coincides with a light-sheet. Instead, this case gives rise to two nontrivial constraints on two different entropies: one from the GSL and one from the covariant bound. 

Our result allows us to connect a number of older works concerning Bekenstein's bound~\cite{Bek81}. It was shown long ago~\cite{Bou03} that this bound follows from the covariant bound in the weak gravity regime. At the time, a sharp definition of entropy for either bound was lacking~\cite{Bou03a,Bou03b}. A differential definition of entropy was later applied to the right Rindler wedge, and positivity of the relative entropy was shown to reduce to the Bekenstein bound on this differential entropy, in settings where the linear size and the energy of an object are approximately well-defined~\cite{Cas08}. 

Our present work offers two additional routes to the Bekenstein bound, in the sense of providing precise statements that reduce to Bekenstein's bound in the special settings where the entropy, energy, and radius of a system are intuitively well-defined. Combining our result with \cite{Bou03} proves a Bekenstein bound, while supplementing a definition of entropy for both the covariant bound and Bekenstein's bound as the differential entropy on a light-sheet.  The bound is in terms of the product of longitudinal momentum and affine width, but this reduces to the standard form $2\pi ER/\hbar$, for spherical systems that are well-localized to the light-sheet.  Alternatively, we may regard our Sec.~\ref{sec-SK} alone as a direct proof of Bekenstein's bound. Again the bound is on the differential entropy, but now in terms of the modular energy $\Delta K$ on a finite light-sheet.  For a system of rest energy $E$ that is well localized to the center of a light-sheet of width $2R$ in the rest frame, one has $\Delta K \approx 2\pi ER$, so \cite{Bek81} is recovered.

\paragraph{Extensions}
An interesting problem is the extension of our proof to interacting theories. For interacting theories the quantization of fields on the light front is notoriously tricky. One could still try to define the entropy as the difference in von Neumann entropies for spatial intervals, in the limit where the spatial interval becomes null.  In order to explore the properties of the entropy defined in this way one can consider strongly coupled field theories that have a holographic gravity dual. We have followed the recipe of \cite{Cas13} to obtain the modular Hamiltonian in terms of entropy perturbations.  However, we find that $\Delta S = \Delta K $ holds exactly, and not just to first order in an expansion for states close to the vacuum. That is, the relative entropy for every state is zero. This means that in the light-like limit, the operator algebra on the null interval becomes trivial, and all states on the null interval become indistinguishable.

We expect that this property should extend to interacting theories without a gravity dual.  One can intuitively understand this as follows.  Concentrating on a null interval is equivalent to exploring the theories at large energies, since we want to localize the measurements at $x^-=0$. In an interacting theory this produces parton evolution as in the DGLAP equation \cite{GriLip72, AltPar77, Dok77}. This evolution leads to states that all look the same at high energies. We expect the same equation $\Delta S=\Delta K$ to hold for non-superrenormalizable theories because, in contrast to the free theories we have discussed in this paper, these do not have operators localizable on a finite null surface \cite{Schl72,Ste63}. We plan to discuss these issues further in a separate publication. 

Here we only note that we again find a local form for the modular Hamiltonian for the null surface:
\begin{equation}
  K_L=2\pi \int d^{d-2}x_{\perp}\,\int_0^1 \, dx^+\,\bar{g}(x^+) T_{++}(x^+,x_{\perp})\,.
\end{equation}
Here $\bar{g}(x^+)$ is not given by the same function, $x^+(1-x^+)$, as in the free case (\ref{eq-k}), but it still satisfies all properties stated in the Appendix. Hence the present proof of the covariant bound also applies in this interacting case.   

\begin{figure}[ht!]
\begin{center}
\includegraphics[scale=0.6]{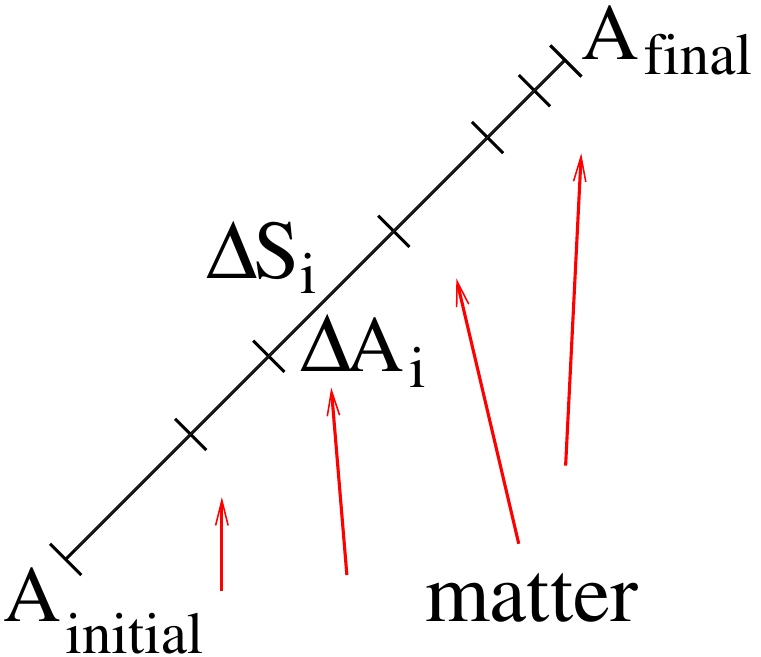}
\caption{A possible approach to defining the entropy on a light-sheet beyond the weak-gravity limit. One divides the light-sheet into pieces which are small compared to the affine distance over which the area changes by a factor of order unity. The entropy is defined as the sum of the differential entropies on each segment.}
\label{splittingnullsurface}
\end{center}
\end{figure}
Another question is how to extend our definition of entropy, and our proof, to the more general situation of a rapidly evolving light-sheet in a general spacetime.  One approach is to divide the light-sheet into small segments along the affine direction in such a way that the change in area is small and then do an approximately flat space analysis for each piece. This is shown in figure \ref{splittingnullsurface}. Here the initial expansion could be large and negative, but this just helps in obeying the bound. Thus, for each segment we  obtain a constraint
 $\Delta A_i/( 4 G \hbar ) \geq \Delta S_i$.  To make this argument we need to have a notion of local vacuum in the QFT in order to define the modular Hamiltonian and to compute $\Delta S$. We assume that this is possible.
Then, for the original region we end up with a bound of the type 
\begin{equation} \label{PossibleBound}
{ \Delta A \over 4 G \hbar }  = { \sum_i \Delta A_i \over 4 G \hbar } \geq \sum_{i} \Delta S_i 
\end{equation}
where $\Delta S_i$ are the entropies differences, as in (\ref{EntropyDifference}), for each of the consecutive null segments. We can take the right hand side of (\ref{PossibleBound}) as the {\it definition} of the total entropy flux.\footnote{We thank D.\ Marolf for this suggestion.} It would be desirable to have a definition of the right hand side which involves the whole null interval. Nevertheless, already 
(\ref{PossibleBound})   is a nontrivial bound. In the regime where we have a clear entropy flux, such as a star or a bucket of water, it reduces to the expected entropy flux if one takes the intervals to be large enough to capture many of the infalling particles.

\bigskip

\acknowledgments 

We thank D.\ Marolf, A.\ Strominger and A.\  Wall for discussions. R.B.\ and Z.F.\ are supported in part by the Berkeley Center for Theoretical Physics, by the National Science Foundation (award numbers 1214644 and 1316783), by the Foundational Questions Institute grant FQXi-RFP3-1323, by ``New Frontiers in Astronomy and Cosmology'', and by the U.S.\ Department of Energy under Contract DE-AC02-05CH11231.  H.C.\ thanks the Institute for Advanced Study for hospitality and financial support. H.C.\ is partially supported by CONICET, CNEA, and Univ. Nac. Cuyo, Argentina. J.M.\ is supported in part by U.S.\ Department of Energy grant DE-SC0009988.

\appendix

\section{Monotonicity of $\frac{\Delta A(c,b)}{4G\hbar}- \Delta S$}
\label{sec-appa}

In Secs.~\ref{sec-SK} and \ref{sec-KA}, we showed that $0 \le \Delta A(c,b)/4G\hbar - \Delta S$. In fact, this difference decreases {\em monotonically} to zero as the boundaries $b$ and $c$ are moved together. To establish this stronger result, it suffices to consider variations of $c$. We may set $b=0$.

We first note that $\Delta K - \Delta S$ is monotonically decreasing when the lightsheet is restricted. This follows immediately from the monotonicity property of relative entropy $S(\rho|\sigma) = \Delta K - \Delta S$ under restriction to a subspace (via a partial trace operation), or more generally under any completely positive trace-preserving map \cite{Lin75}.

Thus it only remains to be shown that $\hbar\delta(c) \equiv \Delta
A(c,0) / 4 G - \Delta K(c,0)$ will decrease monotonically under restriction. We will prove this for the modular Hamiltonian of a free scalar field in first subsection. In the second subsection, with a view to future investigations of the interacting case, we will establish simple sufficient conditions on the modular Hamiltonian from which monotonicity follows.

\subsection{Free scalar field}

Eq.~(\ref{eq-inttheta}) for the area difference and Eq.~(\ref{eq-k}) for the modular Hamiltonian can easily be generalized to an interval of length $c$. Their difference is 
\begin{equation}
  \delta(c) = \int d^2 x_\perp \, \left[-\frac{\theta_0(c)}{4 G}
    + 2\pi\int_0^c d\lambda \, \frac{(c - \lambda)^2}{c}
    T_{kk}(\lambda) \right].
\end{equation}
As we vary $c$, we always choose the initial expansion to be the
largest value compatible with the light-sheet condition:
\begin{equation} \theta_0 = 8 \pi G\inf_{0\le\lambda\le c}
  \int_0^\lambda d\lambda \, T_{kk}(\lambda) \label{eq:theta_b}~.
\end{equation}
The monotonicity of $\delta(c)$ is established by
\begin{equation}
	\frac{d\delta}{dc} = \int d^2 x_\perp \, \left[- \frac{c}{4G} \frac{\partial \theta_0}{\partial c}-\frac{\theta_0}{4 G } + 2\pi\int_0^c d\lambda \, \left(1 - \frac{\lambda^2}{c^2}\right) T_{kk}(\lambda) \right]. \label{eq:partial_delta}
\end{equation} 
The first term is non-negative, since increasing $c$ broadens the range of the $\inf$ in Eq.~(\ref{eq:theta_b}). The latter two terms are together non-negative. This follows from the non-expansion condition by integrating $ \int_0^c d\eta \, \eta \theta(\eta) \le 0$. It follows that $\delta$ is monotonically decreasing under restriction (and monotonically increasing under extension) of the light-sheet. This proves our claim.

\subsection{Sufficient Conditions For Monotonicity}

Now consider a more  general modular Hamiltonian\footnote{As will be discussed in a future publication, 
we expect that an interacting field theory would have a modular Hamiltonian of this type for null intevals. (By contrast, the modular Hamiltonian for spatial regions need not be an integral over local operators.)}
\begin{equation}
\Delta K = \frac{2\pi}{\hbar} \int d^2 x_\perp\int_0^c d\lambda\, g(\lambda,c)\, T_{\lambda\lambda}(\lambda)~.
\end{equation}
We may set $2\pi/\hbar = 4G = 1$ in what follows. Symmetry under time reversal
implies $g(\lambda,c) = g(c-\lambda,c)$, and boost symmetry implies
that
\begin{equation}
g(\lambda,c) = c \bar g(\bar \lambda)~,
\label{eq-gscale}
\end{equation}
where $\bar\lambda = \lambda/c$. We will now show that monotonicity of
$\Delta A-\Delta K$ is guaranteed if $g$ satisfies a small number of other
simple properties of $g$, including concavity.

We have
\begin{equation}
\frac{d\delta}{dc} = -c \frac{d\theta_0}{dc} + \left[-\theta_0 +\int_0^c
  d\lambda 
\left(1-\frac{\partial g}{\partial c}\right) T_{\lambda\lambda}(\lambda)\right] 
\end{equation} 
The first term is nonnegative independently of $g$. The second term is
nonnegative if the function $\partial g/\partial c$ (viewed as a
function of $\lambda$, at fixed $c$) satisfies the following properties:
\begin{eqnarray}
\frac{\partial g}{\partial c} (0) & = & 0~,\\
\frac{\partial g}{\partial c} (1) & = & 1~,\\
\frac{d}{d\lambda}\left(\frac{\partial g}{\partial c}\right) & \geq & 0~.
\end{eqnarray} 
This follows from the nonexpansion condition, via $0\geq \int_0^c
d\lambda\, \theta \frac{d}{d\lambda}(\frac{\partial g}{\partial c}) $.

By Eq.~(\ref{eq-gscale}) we have
\begin{equation}
  \frac{\partial g}{\partial c} = 
  \bar g(\bar\lambda) - \bar\lambda \frac{\partial g}{\partial\bar\lambda}~.
\end{equation}
Hence the above three sufficient conditions for monotonicity are
equivalent to the following conditions
\begin{eqnarray}
\bar g(0) & =& \bar g(1) = 0~,\\
\bar g'(0) &=& -\bar g'(1) = 1~,\\
\bar g'' & \leq  & 0~,
\end{eqnarray} 
where we have also used the symmetry $\bar \lambda \to 1-\bar\lambda$. 

The first two of these conditions are satisfied because the modular
Hamiltonian must reduce to the Rindler Hamiltonian near any
two-dimensional spatial boundary. The last condition is concavity; it
might be related to strong subadditivity. Subject to these conditions,
the GCEB will be satisfied for any state, with monotonically
increasing room to spare as the size of the light-sheet is increased.

\bibliographystyle{utcaps}
\bibliography{all}

\end{document}